\documentclass[twocolumn,english,aps,superscriptaddress,prl]{revtex4-1}

\usepackage{babel}
\usepackage{amsmath}
\usepackage{amssymb}
\usepackage{graphicx}
\usepackage{ytableau}

\usepackage{hyperref}

\usepackage{comment}

\usepackage{epstopdf}
\usepackage{xcolor}
\usepackage{xcolor}
\usepackage{tikz}
\usetikzlibrary{decorations}
\usetikzlibrary{decorations.pathmorphing}
\usetikzlibrary{decorations.pathreplacing}
\usetikzlibrary{decorations.markings}
\usetikzlibrary{shapes.misc}
\usetikzlibrary{calc}

\newcommand{\shsc}[3]{\mathrel{\raisebox{#1}{\protect\scalebox{#2}{#3}}}}

\hypersetup{
    colorlinks=true,
    linkcolor=blue,
    citecolor=blue,    
    urlcolor=blue,
}

\begin{document}

\title{Edge states and universality class of the critical two-box symmetric $\mathrm{SU}(3)$ chain}

\author{Pierre Nataf}
\affiliation{Laboratoire de Physique et Mod\'elisation des Milieux Condens\'es, Universit\'e Grenoble Alpes and CNRS, 25 avenue des Martyrs, 38042 Grenoble, France}
\author{Samuel Gozel}
\affiliation{Institute of Physics, Ecole Polytechnique F\'ed\'erale de Lausanne (EPFL), CH-1015 Lausanne, Switzerland}
\author{Fr\'ed\'eric Mila}
\affiliation{Institute of Physics, Ecole Polytechnique F\'ed\'erale de Lausanne (EPFL), CH-1015 Lausanne, Switzerland}

\date{\today}
\begin{abstract} 
We numerically demonstrate that, although it is critical, the two-box symmetric $\mathrm{SU}(3)$ chain possesses edge states in the adjoint representation whose excitation energy scales with the number of sites $N_s$ as $1/(N_s \log N_s)$, in close analogy to those found in half-integer $\mathrm{SU}(2)$ chains with spin $S\ge 3/2$. We further show that these edge states dominate the entanglement entropy of finite chains, explaining why it has been impossible so far to verify with DMRG simulations the field theory prediction that this model is in the $\mathrm{SU}(3)_1$ universality class. Finally, we show that these edge states are very efficiently screened by attaching adjoint representations at the ends of the chain, leading to an estimate of the central charge consistent within 1\% with the prediction $c=2$ for $\mathrm{SU}(3)_1$.
\end{abstract}

\maketitle


The $\mathrm{SU}(N)$ Heisenberg model is currently attracting a lot of attention because, as soon as $N>2$, not so much is known about its properties while there is a real prospect of implementing it with ultracold fermions~\cite{
wu_exact_2003,
honerkamp_ultrcold_2004,
ultracold_cazalilla_2009,
gorshkov_two_2010,
taie_su6_2012,
pagano_one_2014,
scazza_observation_2014,
zhang_spectroscopic_2014,
hofrichter_direct_2016,
capponi_phases_2016}. 
For the fundamental representation, the $\mathrm{SU}(N)$ model is nothing but a model of quantum permutation between objects with $N$ colors, and in one dimension, there is a Bethe ansatz solution for any $N$~\cite{sutherland}. The system is critical, with algebraic correlations, and its low-energy, long-range properties are described by a field theory known as the Wess-Zumino-Witten $\mathrm{SU}(N)_1$ universality class, with central charge $c=N-1$ and specific scaling dimensions~\cite{affleck_prl_1986,affleck_critical_1988}. For $\mathrm{SU}(2)$ the fundamental representation corresponds to spin-$1/2$, but it is well known since the work of Haldane that the physics can be very different for other irreducible representations (irreps). For the spin-$1$ case (represented by a horizontal Young diagram with two boxes, $\shsc{0pt}{0.4}{\ydiagram{2}}$), Haldane predicted that the spectrum is actually gapped~\cite{haldane_nonlinear_1983,haldane_continuum_1983}, a result confirmed soon after numerically~\cite{white_density_1992,white_numerical_1993} and experimentally~\cite{buyers_1986}.

It is then a very natural question to wonder whether and how these results can be generalized to $N>2$~\cite{affleck_exact_1986,
affleck_critical_1988,
greiter_valence_2007,
rachel_spinon_2009,
bykov_geometry_2013,
lajko_generalization_2017,
tanizaki_anomaly_2018,
wamer_selfconjugate_2019,
wamer_generalization_2020,
wamer_mass_2020}. In order of increasing complexity, the next case is the $\mathrm{SU}(3)$ model in the symmetric representation with two boxes, the same Young diagram $\shsc{0pt}{0.4}{\ydiagram{2}}$ as the spin-$1$ chain for $\mathrm{SU}(2)$. It turns out that this model has not yet received a compelling solution, by which we mean, in the absence of exact results, a field theory prediction confirmed by numerical results. According to field theory, if the system is critical, it can only be in the $\mathrm{SU}(3)_1$ universality class because the only alternative, $\mathrm{SU}(3)_2$, has a relevant operator allowed by symmetry~\cite{lecheminant_massless_2015}. And a generalization of Haldane's semiclassical argument has suggested that the system is indeed critical because the only cases where there is no topological term in the action, hence where the system must be gapped, are those where the number of boxes is a multiple of three~\cite{lajko_generalization_2017} (a result recently confirmed numerically for three boxes~\cite{gozel_2020}). However, it has proven impossible so far to confirm that the two-box symmetric $\mathrm{SU}(3)$ chain is in the $\mathrm{SU}(3)_1$ universality class. Results of exact diagonalizations on small chains with periodic boundary conditions are rather consistent with $\mathrm{SU}(3)_2$~\cite{nataf_exact_2016}, a result interpreted, in view of the field theory prediction, as an evidence that a cross-over must take place as a function of the size to the $\mathrm{SU}(3)_1$ universality class. In these circumstances, the obvious thing to do is to try density-matrix renormalization group (DMRG) calculations~\cite{white_density_1992,
white_density_1993} which, with open boundary conditions, can be performed on chains with hundreds of sites. However, much to our surprise, the central charge deduced from fitting the entanglement entropy with the Calabrese-Cardy formula~\cite{calabrese_entanglement_2004} leads to a result much larger than the expected $c=2$ (see below), a difficulty also met by other groups~\cite{private}. So as of today the two-box symmetric $\mathrm{SU}(3)$ chain remains a puzzle.

In this Letter, we identify the origin of the problem as being due to edge states. In gapped topological phases such as the spin-$1$ chain, it is well known that edge states can exist, and in open chains they show up as low-lying excitations inside the gap~\cite{Kennedy_1990, schollwock_haldane_1995}. In gapless systems, this is less well known, but it has been shown in the nineties that the $S=3/2$ chain has edge states that lead to an excitation with an energy smaller than the finite-size gap by a factor $\log N_s$, where $N_s$ is the number of sites~\cite{Ng_1994,Ng_1995,gabor_2006}. More precisely, we will show that the two-box symmetric $\mathrm{SU}(3)$ chain also has such an excitation below the finite-size gap due to edge states, that it disappears when we screen the edge states by attaching adjoint representations at the ends of the chain, and that the entanglement spectrum then leads with excellent accuracy to the expected central charge $c=2$ when fitted with the Calabrese-Cardy formula.

The Hamiltonian of the $\mathrm{SU}(3)$ Heisenberg chain can be written quite generally in terms of the $\mathrm{SU}(3)$ generators as
\begin{equation}
    \label{eq:su3-hw-H}
    \mathcal{H} = J \sum\limits_{i} \sum\limits_{\alpha,\beta=1}^{3} \mathcal{S}^{\alpha\beta}_i \mathcal{S}^{\beta\alpha}_{i+1}.
\end{equation}
For symmetric irreps, and up to a constant, this can be rewritten in terms of boson creation and annihilation operators as
\begin{equation}
    \label{eq:su3-hw-H}
    \mathcal{H} = J \sum\limits_{i} \sum\limits_{\alpha,\beta=1}^{3} b^{\alpha\dagger}_i b^{\beta}_i b^{\beta\dagger}_{i+1} b^{\alpha}_{i+1}.
\end{equation} 
In the following, we will concentrate on the case with two bosons per site, which corresponds to the $6$-dimensional symmetric irrep described by the Young diagram $\shsc{0pt}{0.4}{\ydiagram{2}}$. The $6$ states correspond to all the ways of constructing a two-boson state with three colors. Throughout, we will also use the alternative notation [$\alpha _1$,$\alpha _2$,$\alpha _3$] for the irreps of $\mathrm{SU}(3)$, where the integers $\alpha _1$, $\alpha _2$ and $\alpha _3$ correspond to the lengths of the rows in the corresponding Young diagram \footnote{Irreps of SU(N) are usually described with $N-1$ integers corresponding, for instance, to the lengths of the rows of the Young diagram, as any column with $N$ boxes corresponds to a totally antisymmetric combination of particles and can thus be removed. Here, however, we use a notation with $N=3$ integers for the irreps of $\mathrm{SU}(3)$ because, in practice, we map irreps of $\mathrm{SU}(N)$ with $n$ boxes onto irreps of the permutation group $S_n$ with at most $N$ rows. See Refs.~\cite{nataf_exact_2014,gozel_2020}  for details.}. With these notations, the two-box symmetric irrep is denoted by [2,0,0].

To reach long enough chains, we will use the same version of DMRG as that used previously for the fundamental representation~\cite{nataf_density_2018} and for the three-box symmetric representation~\cite{gozel_2020}. It relies on a basis constructed with the help of standard Young tableaus~\cite{young_onsubstitutional_1900,
rutherford_substitutional,
nataf_exact_2014,
nataf_exact_2016,
wan_exact_2017}, and allows one to diagonalize the Hamiltonian directly in arbitrary irreps of $\mathrm{SU}(3)$ except those that are degenerate when making the product of two half chains~\cite{outer_multiplicity}, a limitation with no major impact for the problem addressed here. 

\begin{figure}[h!]
\includegraphics[width=0.48\textwidth]{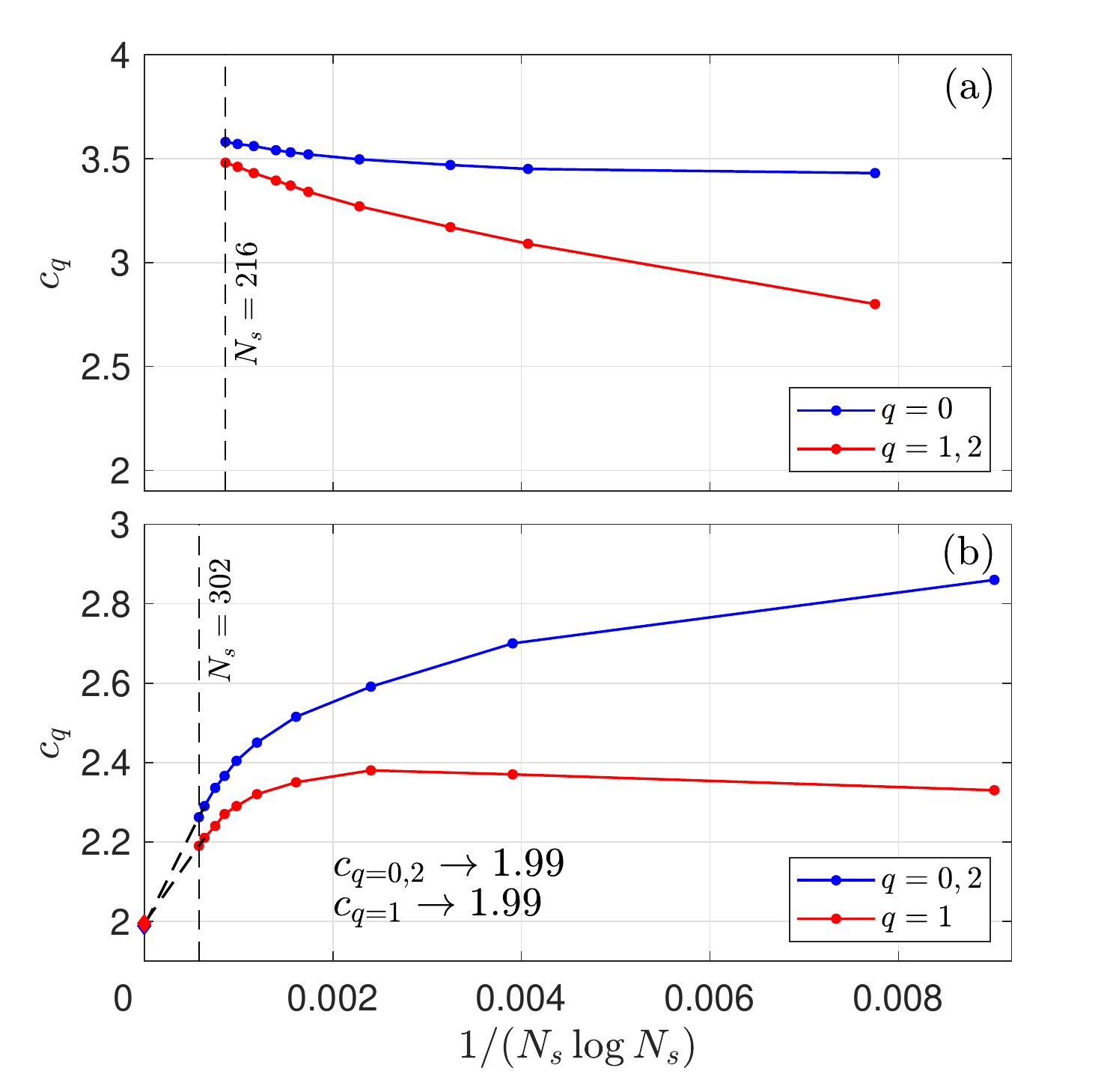}
\caption{Central charges extracted from the entanglement entropy for the Heisenberg open chain of size $N_s$ with the two-box symmetric irrep at each site. In  the top panel (a),  it increases with $N_s$ to values larger than $3.5$. There are two sets of points ($q=0$ and $q=1,2$) because the entanglement entropy oscillates between a top and a bottom envelope (see~\cite{supp_mat}). For the bottom panel (b), we have added edge spins living in the adjoint irrep which screen the edge states, and the central charges converge towards a value very close to $2$, in agreement with field theory arguments. The results have been plotted as $1/(N_s \log N_s)$ by analogy with the edge state gap. See text for details.}
\label{fig1}
\end{figure}

As a first attempt at characterizing the universality class of that model, we have performed DMRG simulations on open chains, and we have fitted the entanglement entropy with the Calabrese-Cardy formula \cite{calabrese_entanglement_2004} to extract the central charge. The finite-bond dimension effects on the central charge are small and have been taken care of by scaling the results with the discarded weight (see Supplemental Material~\cite{supp_mat} for an example). The results are plotted in the top panel of Fig.~\ref{fig1}. Because of period $3$ oscillations, it is better to fit independently two sets of points ($q=0$ and $q=1,2$, where $q$ is the position of the cut modulo $3$ for the calculation of the entanglement entropy along the chain~\cite{supp_mat}). In both cases, the apparent central charge increases upon increasing the system size to values of the order of $3.6$ or $3.7$, well above $16/5$, the value for $\mathrm{SU}(3)_2$ observed on small chains with periodic boundary conditions~\cite{nataf_exact_2016}, and a fortiori much larger than the expected $c=2$ for $\mathrm{SU}(3)_1$. This result makes no physical sense, and the only possibility is that the entanglement we are measuring is not that of the bulk, but is dominated by edge effects.
So let us have a closer look at the spectrum of open chains. 

\begin{figure}[h!]
\includegraphics[width=0.48\textwidth]{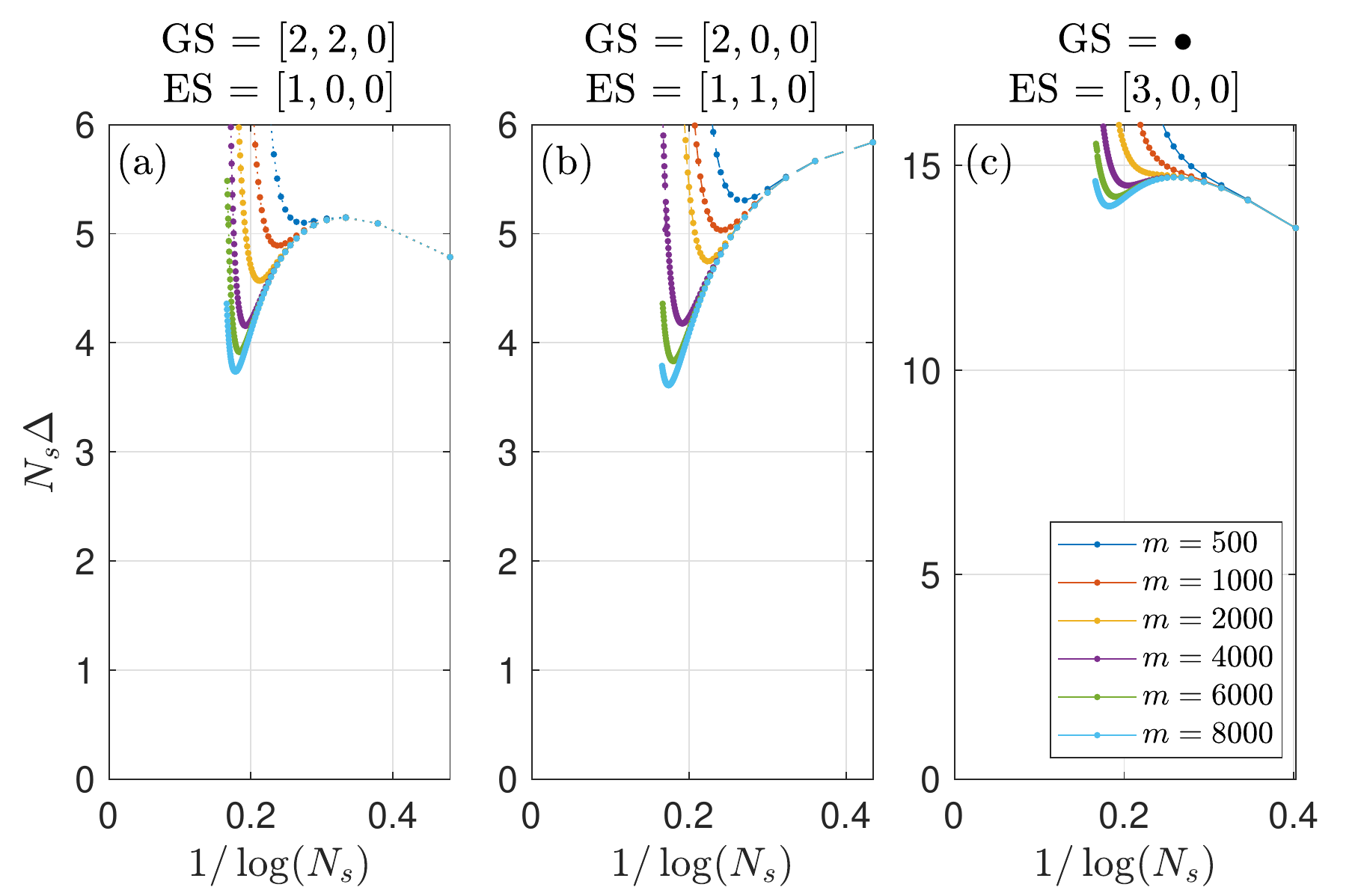}
\caption{$N_s \Delta$ for (a) $N_s=1$ mod. $3$, (b) $N_s=2$ mod. $3$ and (c) $N_s=0$ mod. $3$, where $\Delta$ is a gap between the ground state (GS) energy and the first excited state (ES) energy living in the irrep (a) $[1,0,0]$ for $N_s=1$ mod. $3$, (b) $[1,1,0]$  for $N_s=2$ mod. $3$ and (c) $[3,0,0]$ for $N_s=0$ mod. $3$.
The results are shown as a function of $1/\log N_s$ for different values of $m$, i.e. the number of states kept controlling the DMRG accuracy. 
For $N_s=1,2$ mod. $3$, $N_s \Delta$ goes to zero when $N_s \rightarrow \infty$, implying that, unlike the usual expectation for a bulk excitation in a gapless chain, the excitation shown does not vanish as $1/N_s$ but faster, revealing the presence of edge states. By contrast, when $N_s=0$ mod. $3$, $N_s \Delta$ goes to a strictly positive constant when $N_s \rightarrow \infty$ because the sector is not that of the lowest excitation (See text for details).}\label{fig2}
\end{figure}

Since we are dealing with an irrep with two boxes at each site, the ground state can only be a singlet if the number of sites $N_s$ is a multiple of $3$. If the number of sites is equal to $1$ mod. $3$, the ground state is expected to be in the [2,0,0] sector, and if it is equal to $2$ mod. $3$, the ground state should be in the [2,2,0] sector. Quite generally, the first excited state is expected to be in the adjoint representation [2,1,0] if the ground state is a singlet, and in the most antisymmetric combination of the adjoint with the ground state sector otherwise, i.e. [1,1,0] if the ground state is in the [2,0,0] sector, and [1,0,0] if the ground state is in the [2,2,0] sector. Since we cannot calculate in the adjoint sector for technical reasons, we have calculated the first excitation in the [3,0,0] irrep for $N_s$ multiple of three, and the expected lowest sector in the other cases. The results are shown in Fig.~\ref{fig2}. If the spectrum was simply representative of the bulk, the first excitation should collapse onto the ground state as $1/N_s$ if the system is gapless, and $N_s$ times the energy should go to a constant. This is clearly not the case when $N_s$ is not a multiple of three: $N_s$ times the excitation energy goes to zero as $1/\log N_s$. This is typical of edge states in half-odd-integer spin chains with $S\ge 3/2$~\cite{Ng_1994,Ng_1995,Lou_2000,gabor_2006}, and by analogy with that case, we conclude that there are edge states that produce an excitation in the adjoint representation with an energy scaling as $1/(N_s \log N_s)$. By contrast, when $N_s$ is a multiple of $3$, the first sector we can target, [3,0,0], is not that of the first excitation, and the corresponding excitation scales as $1/N_s$, a scaling typical of elementary bulk excitations in 1D gapless systems and in sharp contrast with the previous edge excitations calculated for $N_s$ not a multiple of $3$.

In the spin-$3/2$ chain, a simple argument to understand the presence of spin-$1/2$ edge states is to see it as a gapless spin-$1/2$ chain coupled ferromagnetically to a gapped spin-$1$ chain, i.e. a ladder with ferromagnetic rungs and with antiferromagnetic spin-$1/2$ and spin-$1$ legs respectively. The spin-$1/2$ chain has no edge states, but the spin-$1$ chain has spin-$1/2$ edge states, and the resulting picture is that of a spin-$1/2$ chain with spin-$1/2$ edge states coupled ferromagnetically to it, leading to the equivalent of the ferromagnetic Kondo problem and to the $1/(N_s \log N_s)$ scaling of the lowest excitation~\cite{eggert_1992}. By analogy, we can see the on-site two-box symmetric irrep as obtained from the tensor product of a three-box symmetric irrep and a two-box antisymmetric irrep. Indeed
\begin{equation}
\shsc{0pt}{0.6}{\ydiagram{3}} \otimes \shsc{3pt}{0.6}{\ydiagram{1,1}} \ = \ \shsc{3pt}{0.6}{\ydiagram{4,1}} \ \oplus \ \shsc{0pt}{0.6}{\ydiagram{2}}.
\end{equation}
The two-box antisymmetric irrep is gapless and described by $\mathrm{SU}(3)_1$ since $\shsc{2pt}{0.4}{\ydiagram{1,1}}$ is the complex conjugate irrep of the fundamental irrep $\shsc{0pt}{0.4}{\ydiagram{1}}$, while the irrep $\shsc{0pt}{0.4}{\ydiagram{3}}$ is gapped and has adjoint edge states~\cite{gozel_2020}. Note that in that case the coupling between the two chains should be antiferromagnetic to pick the two-box symmetric irrep. In other words, this amount to seeing the two-box symmetric chain as an antiferromagnetic ladder with legs in the two-box antisymmetric and three-box symmetric representations, respectively. 

\begin{figure}[h!]
\includegraphics[width=0.48\textwidth]{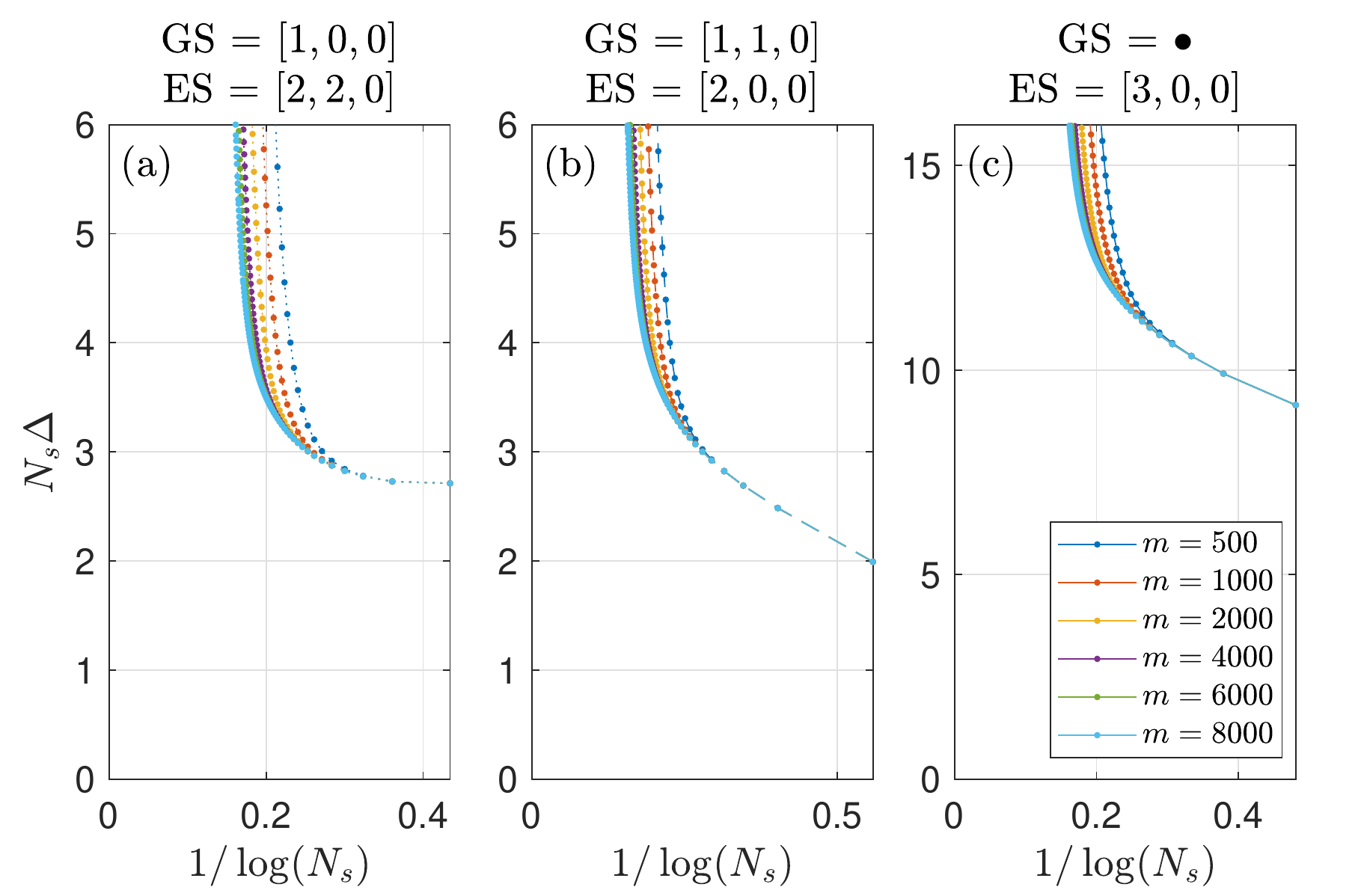}
\caption{Same quantities as in Fig.~\ref{fig2} except that we have added spins living in the adjoint irrep at the two edges of the chain. The calculated gaps $\Delta$ for all the cases considered ($N_s=1,2,0$ mod. $3$ corresponding to left, middle and right panel respectively) vanish as $1/N_s$ (or equivalently $N_s \Delta$ goes to  a strictly positive constant) in the limit $N_s \rightarrow \infty$, by contrast to the situation shown in Fig.~\ref{fig2}, as a consequence of the screening of the edge states by the additional adjoint spins. (See text for details).}
\label{fig3}
\end{figure}

The picture emerging from this analogy is that of a two-box antisymmetric chain with adjoint edge states. Then, if we add edge spins in the adjoint representation coupled antiferromagnetically to the chain, as we did for the three-box symmetric chain \cite{gozel_2020}, we can hope to screen the edge states, and to recover a spectrum typical of a 1D gapless system with excitations scaling as $1/N_s$. As shown in Fig.~\ref{fig3}, this is precisely what happens. In fact, $N_s \Delta$ converges now towards a strictly positive constant in the thermodynamic limit. This is fully consistent with the usual behavior of bulk excitations in gapless systems.

Now that we know how to screen the edge states, we can hope to capture the entanglement of the bulk. So we have performed a systematic DMRG investigation of chains with edge spins in the adjoint representation. The resulting entanglement spectrum can again be fitted with the Calabrese-Cardy formula~\cite{calabrese_entanglement_2004}. The results are plotted in the bottom panel of Fig.~\ref{fig1}, and as expected, they are completely different from those obtained without screening the edge states. For small system sizes, the central charge is significantly larger than $2$ for both sets of points, but the finite-size effects are very different.
Indeed, for large enough sizes, the central charge is consistent with $c=2$ within 1\% after extrapolation to $N_s \rightarrow \infty$ (and within 10 \% for the largest system size, $N_s=302$).
Interestingly, for $q=1$, there is a change of behaviour between $N_s=62$ and $N_s=92$ (second and the third point from the right): the central charge first increases before decreasing towards 2. This might be related to the crossover scenario from $\mathrm{SU}(3)_2$ to $\mathrm{SU}(3)_1$ put forward to explain the ED results on small chains with periodic boundary conditions \cite{nataf_exact_2016}, for which the edge states are of course also absent.


To summarize, we have been able to demonstrate numerically with extensive DMRG simulations that the $\mathrm{SU}(3)$ Heisenberg chain with the two-box symmetric irrep at each site is in the Wess-Zumino-Witten $\mathrm{SU}(3)_1$ universality class with central charge $c=2$. This has been made possible by a careful investigation of the excitation spectrum with open boundary conditions that has revealed the presence of edge states living in the adjoint irrep, and by a calculation of the entanglement spectrum after screening these edge states to be able to extract the central charge. 
In the future, it would be interesting to try and go beyond the simple picture provided above to explain the presence of edge states, and to develop the equivalent of the ferromagnetic Kondo theory for $\mathrm{SU}(3)$. 
At the technical level, it would be useful to improve our numerical machinery in order to be able to target all $\mathrm{SU}(N)$ irreps and to access all the low lying excitation gaps, and to try and calculate the central charge for larger chains with periodic boundary conditions to check for the crossover scenario. The other cases relevant for the generalization of the Haldane conjecture, like the $\mathrm{SU}(4)$ Heisenberg chain with the $p$-box symmetric irrep at each site with $p=2$ and $p=4$, would also be interesting (but challenging) to address numerically~\cite{lajko_generalization_2017}. 

Finally, let us comment briefly on the possible physical implementations of the current model. $\mathrm{SU}(3)$ spins living in any two-column irrep can be simulated using alkaline-earth atoms such as $^{87}{\mathrm Sr}$ or  $^{173}{\mathrm Yb}$~\cite{gorshkov_two_2010,
desalvo_degenerate_2010,
taie_realization_2010,
taie_su6_2012}, and the criticality of the chain can be revealed through the measurement of the two-site correlations,
as experimentally achieved very recently for $\mathrm{SU}(N)$~\cite{taie2020observation}.
Alternatively, quantum interference~\cite{islam_2015,
kaufman_2016} and randomized measurement~\cite{brydges_2019,
elben_2020} based protocols can allow experimentalists to directly access the entanglement properties of many-body cold atoms systems.

\section{Acknowledgments}

We acknowledge very useful discussions with Ian Affleck and Philippe Lecheminant.
This work has been supported by the Swiss National Science Foundation.

\bibliographystyle{apsrev4-1}
\bibliography{references}

\end{document}


\title{Supplemental Material \\ Edge states and universality class of the critical two-box symmetric $\mathrm{SU}(3)$ chain}
\author{Pierre Nataf}
\affiliation{Laboratoire de Physique et Mod\'elisation des Milieux Condens\'es, Universit\'e Grenoble Alpes and CNRS, 25 avenue des Martyrs, 38042 Grenoble, France}
\author{Samuel Gozel}
\affiliation{Institute of Physics, Ecole Polytechnique F\'ed\'erale de Lausanne (EPFL), CH-1015 Lausanne, Switzerland}
\author{Fr\'ed\'eric Mila}
\affiliation{Institute of Physics, Ecole Polytechnique F\'ed\'erale de Lausanne (EPFL), CH-1015 Lausanne, Switzerland}
\date{\today}

\maketitle

\makeatletter
\renewcommand{\theequation}{S\arabic{equation}}
\renewcommand{\thefigure}{S\arabic{figure}}
\renewcommand{\bibnumfmt}[1]{[S#1]}
\renewcommand{\citenumfont}[1]{S#1}

 The analysis of the entanglement entropy and the extraction of the central charge are presented in Section~\ref{sec::EE_without} for the Heisenberg open chain with two-box symmetric irrep at each site and in Section~\ref{sec::EE_with} for the same chain with additional edge spins living in the adjoint irreps.

\section{Entanglement entropy and central charge without edge spins}
\label{sec::EE_without}

\begin{figure}
\begin{center}
\includegraphics[width=0.5\textwidth]{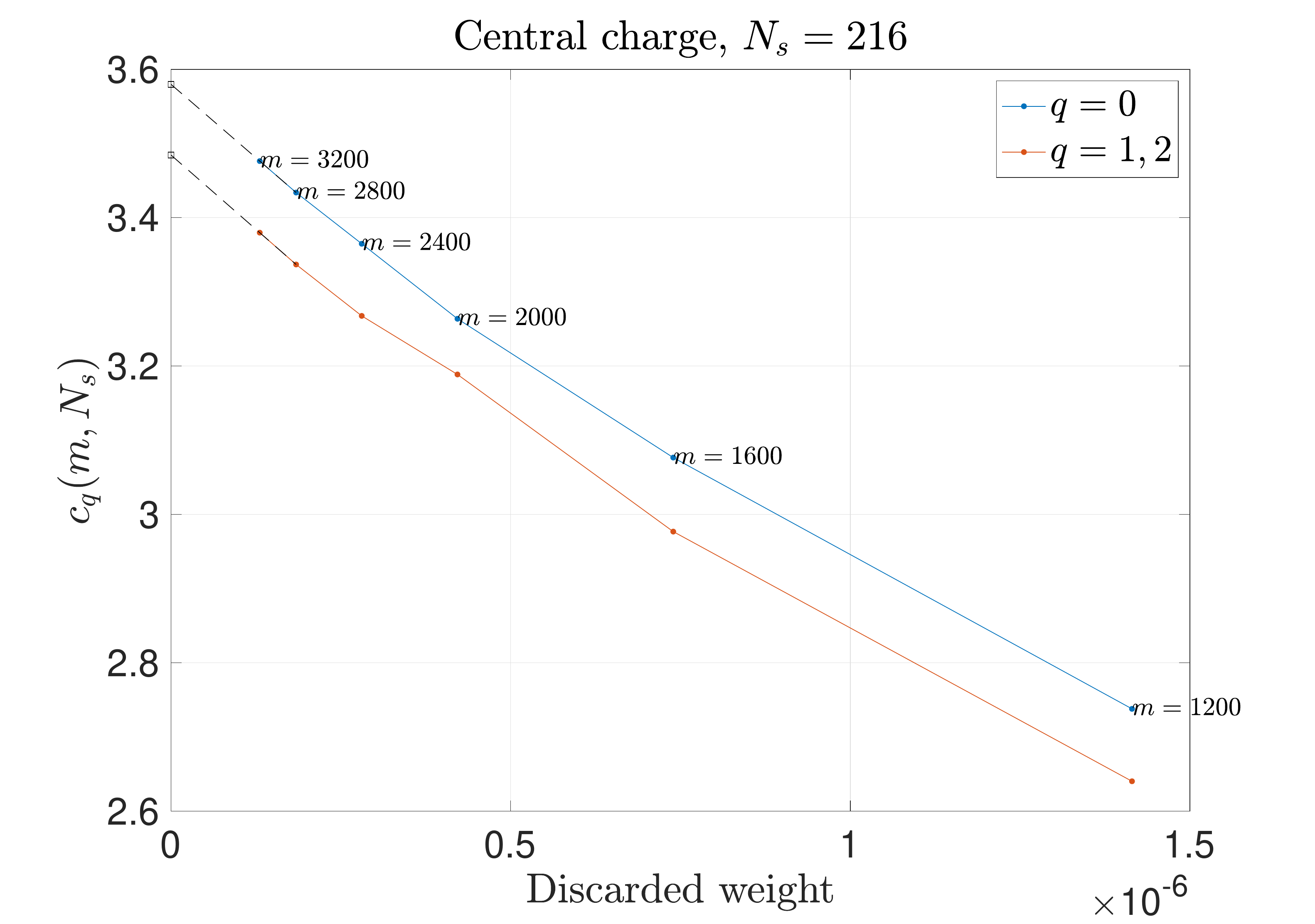}
\end{center}
\caption{Entanglement entropy of a chain with $N_s=216$ sites as a function of the discarded weight for different number of states kept $m$. The extrapolation (for the discarded weight going to $0$ or equivalently $m \rightarrow \infty$) appearing here is then reported in Fig.~1 of the main text.}
\label{fig::extrapol_CC}
\end{figure}

Given the curvature of the entanglement entropy for small chains we fit the Calabrese-Cardy formula~\cite{calabrese_entanglement_2004},
\be
 \label{calabrese_cardy}
S(x) = \frac{c}{6} \ln\left( \frac{2 N_s}{\pi} \sin\left(\frac{\pi x}{N_s} \right) \right) + c_1
\ee
where $c$ is the central charge (for a critical model) and $c_1$ is a non-universal constant term, to our DMRG results as follows. For each value of the number of states kept $m$ we perform the fit on six (for small chains) or eight (for longer chains) points separately for $q=0,1,2$ because of the oscillations appearing in the entanglement entropy along the chain. This defines the ``central charge'' $c_q(m,N_s)$ for finite bond dimension $m$ and finite $N_s$.
It appears that $c_1(m,N_s)=c_2(m,N_s)$ within 1 \% so that we take the average that we call $c_{1,2}(m,N_s)$  to simplify the discussion. 
 We then perform an extrapolation with respect to the discarded weight to obtain $c_q(m=\infty,N_s), \ q=0,1,2$ as shown in Fig.~\ref{fig::extrapol_CC}. 

\begin{figure*}[ht]
\centering
\begin{center}
\includegraphics[width=1\textwidth]{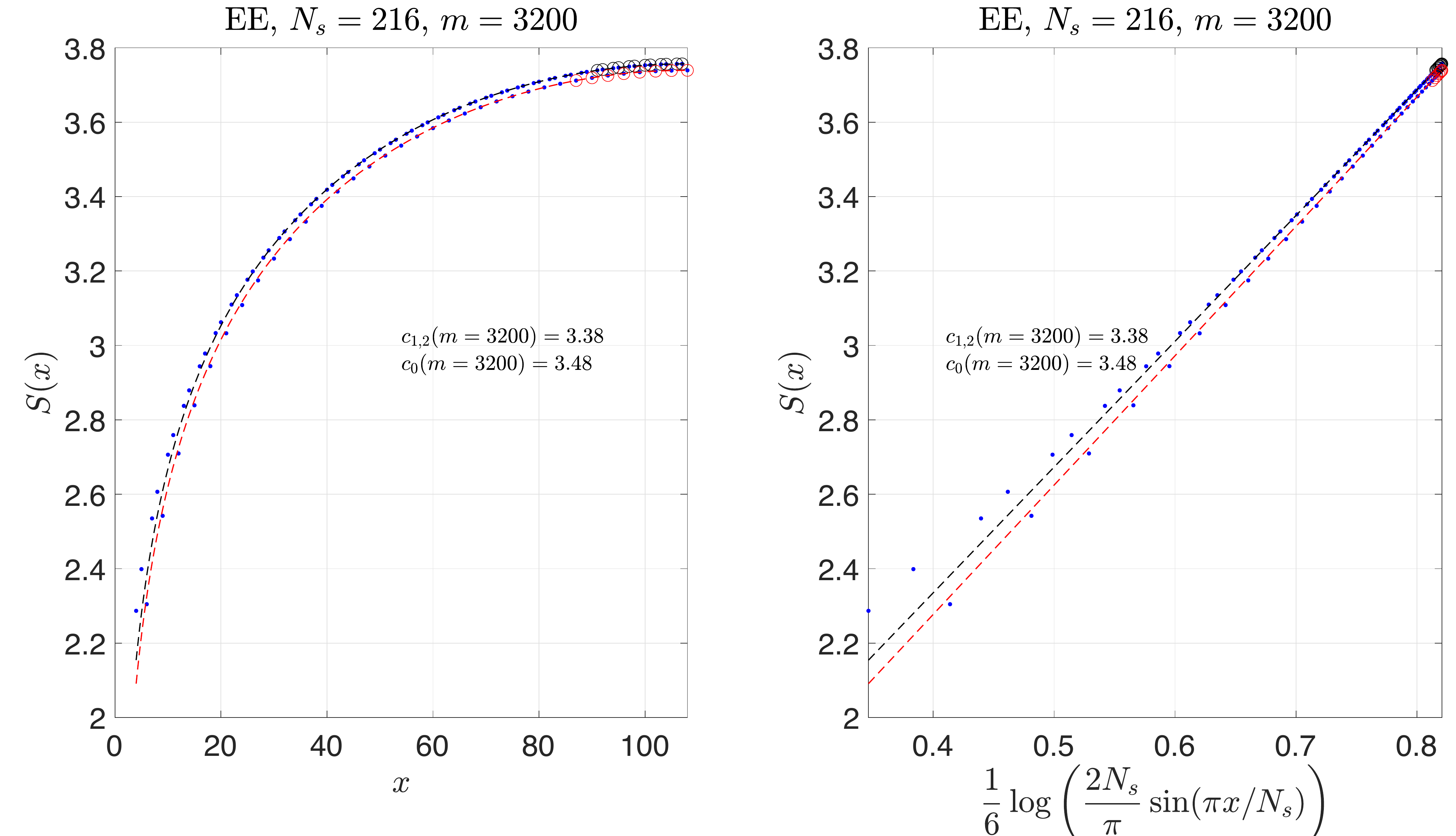}
\end{center}
\caption{Left: Entanglement entropy of an open chain with $N_s=216$ sites and $m=3200$ (number of states kept in the DMRG truncation) as a function of $x$, the location of the cut along the chain. Depending on the value of $q=x$  mod. $3$, two envelopes appear: one for $q=0$ and one for $q=1,2$. Right: same quantity in log scale to extract the central charge through the Calabrese-Cardy formula \eqref{calabrese_cardy}. We have used $8$ points (shown as empty circles here) in the middle of the chain to perform the fit.}
\label{fig::supp:EE}
\end{figure*}

\newpage

\section{Entanglement entropy and central charge with adjoint edge spins}
\label{sec::EE_with}
We proceed here to the very same routine and illustrate the methodology for $N_s=302$ below.

\begin{figure*}[ht]
\centering
\begin{center}
\includegraphics[width=1\textwidth]{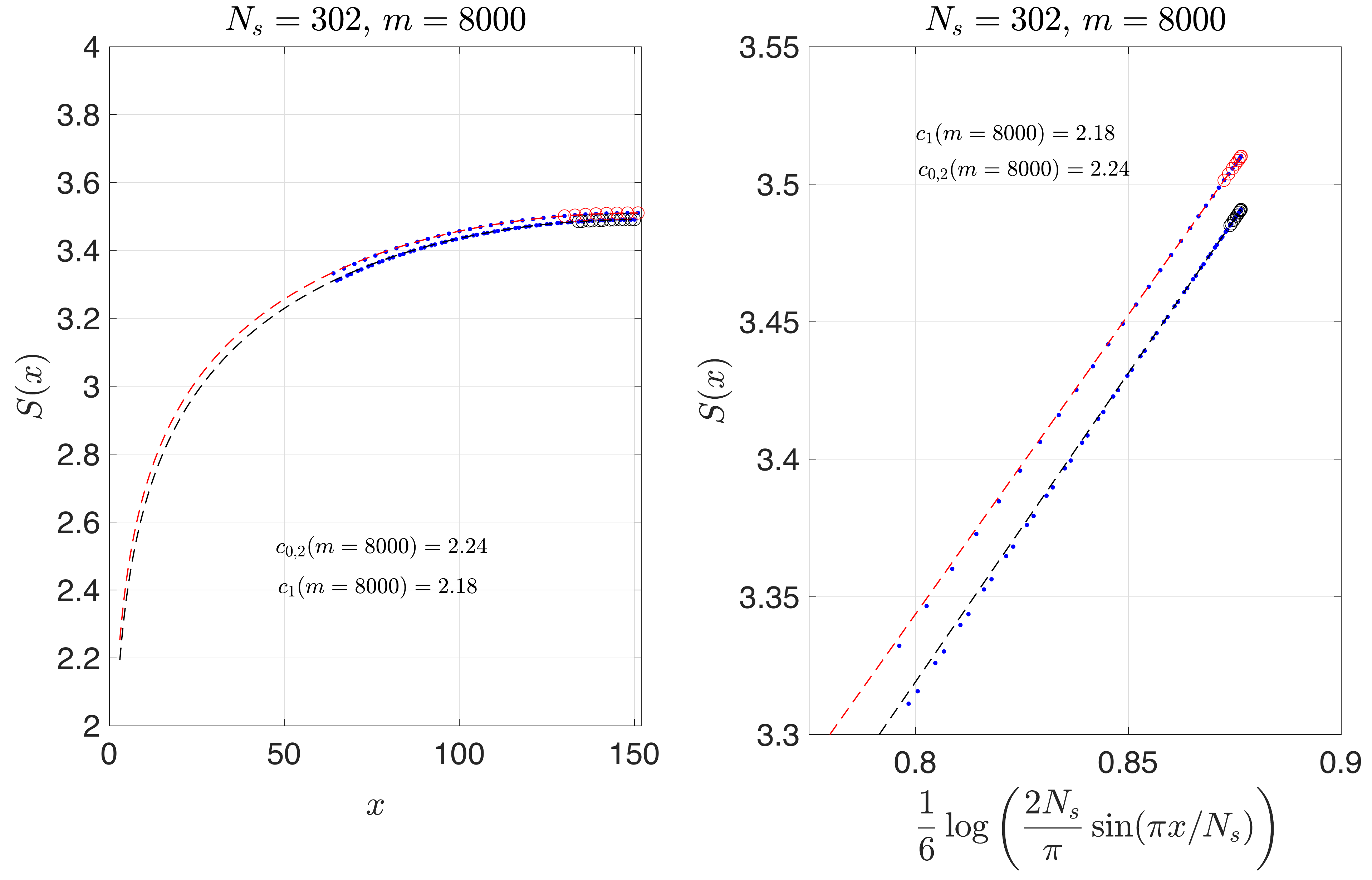}
\end{center}
\caption{Left: Entanglement entropy of an open chain with adjoint edge spins and with $N_s=302$ sites and $m=8000$ (number of states kept in the DMRG truncation) as a function of $x$, the location of the cut along the chain. Depending on the value of $q=x$  mod. $3$, two enveloppes appear: one for $q=0,2$ and one for $q=1$. Right: same quantity in log scale to extract the central charge through the Calabrese-Cardy formula \eqref{calabrese_cardy}. We have used $8$ points (shown as empty circles here) in the middle of the chain to perform the fit.}
\end{figure*}

\begin{figure}
\begin{center}
\includegraphics[width=0.5\textwidth]{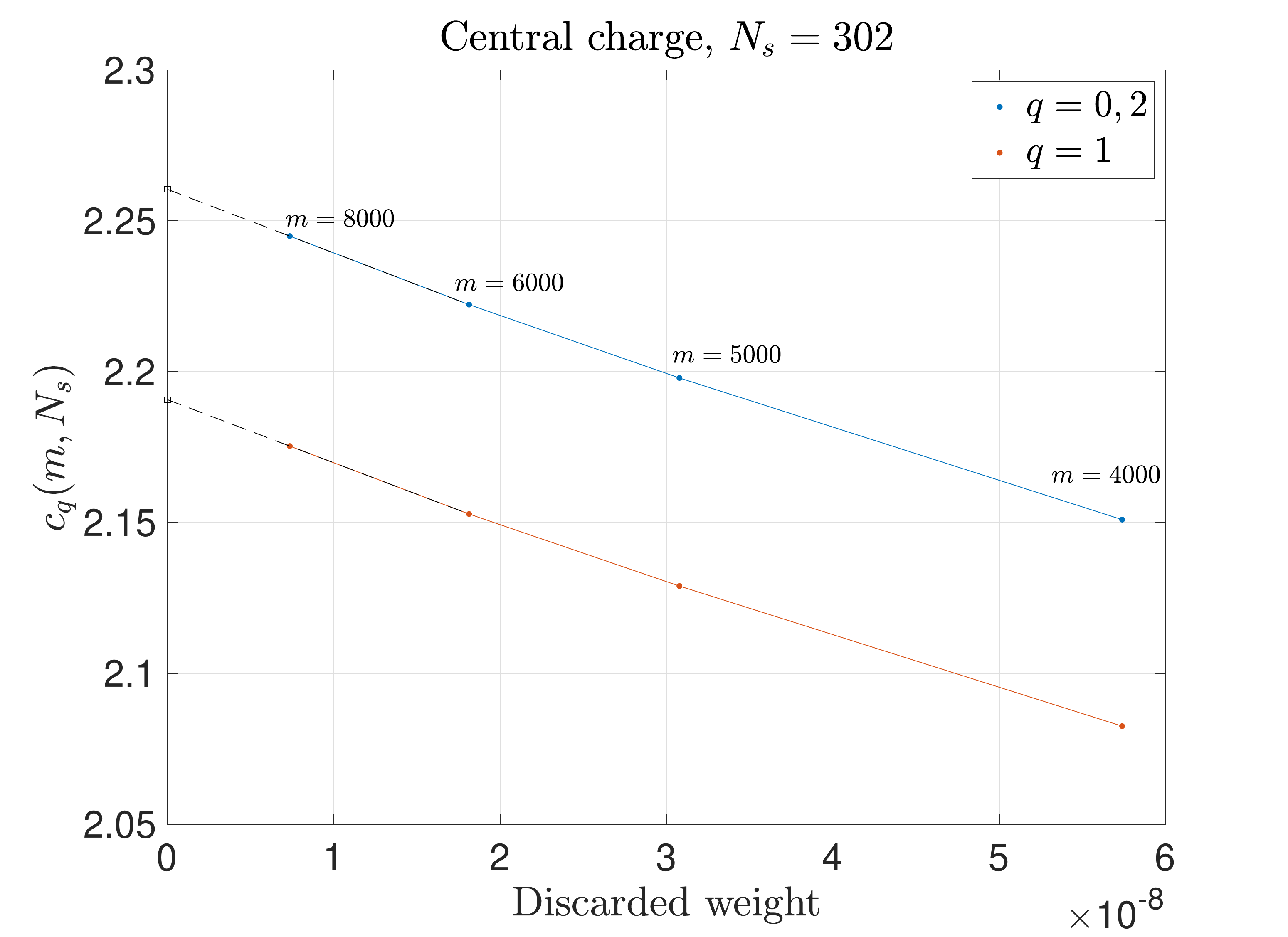}
\end{center}
\caption{Entanglement entropy of a chain with $N_s=302$ sites as a function of the discarded weight for different number of states kept $m$. The extrapolation (for the discarded weight going to $0$ or equivalently $m \rightarrow \infty$) appearing here is then reported in Fig.~1 of the main text.}
\end{figure}

\nocite{*}
\bibliographystyle{apsrev4-1}
\bibliography{references_supp}